\documentclass{ws-p8-50x6-00}

\newcommand{\beq}{\begin{equation}}
\newcommand{\eeq}{\end{equation}}

\begin{document}

\title{Bubbles from Dark Energy ?}

\author{Jochen Weller\footnote{On leave from the Theoretical Physics Group, Blackett Laboratory, Imperial College, Prince Consort Road, London, SW7 2BZ, U.K.}}

\address{Department of Physics, University of California at Davis, Davis, CA 95616,
USA\\E-mail: weller@physics.ucdavis.edu}


\maketitle

\abstracts{In a model recently proposed by Albrecht and
Skordis\cite{Albrecht:99} it was suggested that the observed accelerated
expansion of the universe could be caused by a scalar field which is trapped
in a local minimum of an exponential potential modified by a
polynomial prefactor. We show that scalar
field cosmologies with this kind of local minimum in the potential are stable
and do not decay to the true vacuum if they fulfill the observational
constraints from the Type Ia Supernovae experiments.
Further we briefly sketch how this potential could be related to a potential of
interacting D-branes.} 
\section{Introduction}\label{intro}
Recent observations by the Supernovae Cosmology Project
(SCP)\cite{Perlmutter:98} and the High-Z Supernovae Search Team\cite{Riess:98}
have revealed that there is an energy component in the universe which is dark
and has negative pressure. This is confirmed if one combines
observations of the anisotropy in  
the cosmic microwave background (CMB) radiation and clusters\cite{Knox:00}. The
simplest way to achieve accelerated expansion of the universe is by
introducing an {\em ad hoc} cosmological constant. The findings of the SCP are
that the probability for a non-vanishing cosmological constant is
99\%\cite{Perlmutter:98}. If one assumes a flat universe, which seems
to be confirmed by recent CMB observations, the universe consists of
30\% matter and 70\% cosmological constant or dark energy
component\cite{Knox:00}. A more general way to obtain accelerated
expansion of the universe is by introducing a scalar field $\phi$
which either slowly rolls down a potential or is trapped in a local
minimum. In this way the universe eventually becomes vacuum dominated
and therefore the expansion accelerates. During recent years
dark energy or quintessence models have been
proposed\cite{models:88-00}, with some of these models
needing tuning of the initial conditions to fulfill the observational
constraints, but the majority requiring only a tuning of the parameters
of the model. The novel feature of the model by Albrecht and Skordis\cite{Albrecht:99} is that the parameters involved are roughly of
order one in Planck units ($M_{\rm Pl} \approx 2.44 \times 10^{18}
{\rm GeV}$) and only moderate tuning is required to be within
the observational constraints. Throughout this paper we use natural
units ($\hbar = c = 1$) and set the Planck
mass $M_{\rm Pl} = \left(\hbar c^5/ 8\pi G\right)^{1/2} = 1$.
\section{The dark energy potential and tunneling}\label{main}
\begin{figure}[!h]
\epsfxsize=14pc 
\center{\epsfbox{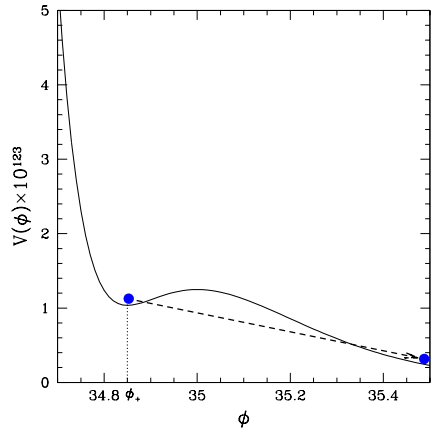}} 
\caption{An example of an exponential potential with a polynomial
prefactor. The parameters are $\alpha=2$, $\beta=34.8$, $\lambda = 8$ and
$\delta =0.01$.   \label{fig:potential}} 
\end{figure}

The dark energy field in the Albrecht and Skordis model rolls down
an exponential potential which is modified by a polynomial
prefactor,
\beq
V(\phi) = V_p(\phi) {\rm e}^{-\lambda \phi}\, , \qquad V_p(\phi) = \left(\phi - \beta\right)^\alpha + \delta\, .
\eeq 
This prefactor leads to a local minimum in the potential in which the
scalar field gets trapped after rolling down the exponential
branch. The initial potential energy of the field is ``redshifted
away'' due to the friction term in the field
equations, caused by the expansion. Therefore the field gets
trapped in the minimum and the expansion starts accelerating. The model was
tested for a range of parameters\cite{Albrecht:99}. To illustrate our
findings with numbers we took the parameters from Albrecht and Skordis\cite{Albrecht:99} with $\alpha=2$, $\lambda=8$ and $\delta = 0.01$.
The observational constraints are that the universe today consists of
30\% matter, 70\% dark energy and we choose the Hubble constant to be $65\; {\rm km}\,{\rm
s}^{-1}\,{\rm Mpc}^{-1}$. To fulfill the observational constraints it
is necessary to adjust the position of the local minimum by tuning
$\beta$ to  $33.989$. The potential shown in Fig.~\ref{fig:potential} is for $\beta=34.8$.
To calculate the transition rate from the false vacuum through the barrier we follow the prescription of Coleman and De Luccia\cite{Coleman:77} . The 
tunneling rate $\Gamma$ per volume element $V$ in the semi-classical
approximation is $\Gamma/V \equiv A{\rm e}^{-B}$. In order to calculate $B$
in the vacuum decay amplitude we need to compute
\beq
	B = S_{\rm E}\left(\phi_{\rm cl}\right)-S_{\rm
	E}\left(\phi_+\right) \, ,
\eeq
with $S_{\rm E}$ being the analytic continuation of the Minkowskian action
\beq
	S = \int
d^4x\sqrt{-g}\left[\frac{1}{2}g^{\mu\nu}\partial_\mu\phi\partial_\nu\phi-V(\phi)-\frac{R}{2}\right]\, ,
\eeq
where $g_{\mu\nu}$ is the metric, $R$ the corresponding curvature
scalar and $\phi_+$ is the field in the local minimum of the potential. 
In order to obtain the non-trivial classical solution $\phi_{\rm cl}$ to the Euclidean
field equations we assume that an $O(4)$ symmetric ansatz minimizes
the Euclidean action $S_{\rm E}$, with the corresponding Euclidean metric of the
form $ds^2 = d\xi^2+\rho(\xi)^2d\Omega^2$, $\xi$ the radial coordinate
and $\rho=\sqrt{t^2+|{\bf x}|^2}$. The boundary conditions\cite{Jensen:84} for the	
tunneling solution in this coordinates are $\phi(\infty) = \phi_+$, $\phi^\prime(\infty) = 0$, $\phi^\prime(0) = 0$ and $\rho(0)
= 0$. We solved the Euclidean field equations under these boundary conditions
numerically and computed the tunneling suppression factor $B$. With the
parameters above we got
\[
	B  \approx 10^{123} \; , 
\] 
where the field reappearing after the barrier at $\phi \approx 35.5$
with $\phi^\prime = 0$, from where it rolls towards $\phi = \infty$,
the {\em true} vacuum state as sketched in Fig.~\ref{fig:potential}. 
The probability for a tunneling event per unit time, per unit volume is 
$\Gamma/V \approx \exp\left(-10^{123}\right)$ which is
negligible. Therefore {\em no} bubbles of true vacuum are formed. In
the Coleman and De Luccia prescription the true vacuum state is
reached for finite field values. One might wonder if their result is
also true if the vacuum state is only reached asymptotically. However
we are only interessted in the tunneling rate through the barrier and
not actually when the field ends up in the true vacuum, so our
calculation is valid. We cross	
checked this result by considering a double Mexican Hat potential, $
V(\phi) \propto \phi^2(\phi^2-\alpha^\prime)(\phi^2-\beta^\prime)$, with
the true vacuum state at a finite field value. If we choose the parameters $\alpha^\prime$ and $\beta^\prime$ to get a 
similar barrier width and height as we get for the polynomial
exponential, the tunneling rate is of the same order of magnitude.
\section{Discussion}
The considerations above show that the Albrecht and Skordis model is a
stable solution and the probability that the field decays to the true
vacuum state is negligible. We also observed that if we move the
feature in the potential to smaller field values $\phi \leq 1$ the
tunneling rate is significant and $B \sim {\cal O}(1)$ in Planck
units. However a feature in the potential at such small field values 
could not fulfill the observational constraints and resolve the dark
energy problem, since the field gets trapped in the local
minimum too early. Since the false vacuum state does not decay for the relevant range of parameters, a classical 
description is sufficient and the field stays trapped in the local minimum.
The dark energy model with a polynomial exponential is therefore a
possible solution to the problem of the missing dark energy. One
might wonder if it is possible to connect such a potential to
fundamental physics. A potential with similar features to the one
discussed above is an inverse polynomial exponential of the form
$V(\phi) =
\left[(\phi-\beta)^2+\delta\right]^{-1}\exp(-\lambda\phi)$. It may be possible  to relate this potential to the massive bulk
modes of interacting D3-branes, which involves similar
contributions\cite{Dvali:98}. In this case the false vacuum state is
also very stable with $B \approx 10^{120}$. The question of how realistic a rational
exponential potential which resolves the dark energy problem is
remains open to this point but is under current investigation.\\
{\bf Acknowledgment}\\
We would like to thank A.~Albrecht for fruitful
collaboration\cite{Weller:00} and acknowledge financial
support from the German Academic Exchange Service (DAAD).


\begin{thebibliography}{99}
\bibitem{Albrecht:99} A.~Albrecht and C.~Skordis,
\Journal{\PRL}{84}{2076}{2000}.
\bibitem{Perlmutter:98} S.~Perlmutter et al.,
\Journal{\em ApJ}{483}{565}{1997}, S.~Perlmutter et al.,
\Journal{\em ApJ}{517}{565}{1999}.
\bibitem{Riess:98} A.~Riess et al., \Journal{\em Astron. J}{116}{1009}{1998}.
\bibitem{Knox:00} S.~Dodelson and L.~Knox, {\em astro-ph}
9909454, \Journal{\PRL}{}{to be
published}{2000}.
\bibitem{models:88-00} J.~A.~Frieman et al.,
\Journal{\PRL}{75}{2077}{1995}, P.~Ferreira and M.~Joyce,
\Journal{\PRD}{58}{023503}{1998}, P.~J.~Steinhardt,
L.~Wang and I.~Zlatev, \Journal{\PRD}{59}{123504}{1999},
Ph. Brax and
J.~Martin, \Journal{\PLB}{468}{40}{1999}, T.~Barreiro, E.~J.~Copeland
and N.~J.~Nunes, {\em astro-ph} 9910214, S.~Dodelson,
M.~Kaplinghat and E.~Stewart, {\em astro-ph} 0002360.
\bibitem{Coleman:77} S.~Coleman, \Journal{\PRD}{15}{2929}{1977},
S.~Coleman and F.~De~Luccia, \Journal{\PRD}{21}{3305}{1980}.
\bibitem{Jensen:84} L.~G.~Jensen and P.~J.~Steinhardt,
\Journal{\NPB}{237}{176}{1984}.
\bibitem{Dvali:98} G.~Dvali and S.~H.~Tye,
\Journal{\PLB}{450}{72}{1999}.
\bibitem{Weller:00} J.~Weller and A.~Albrecht in preparation.
\end{thebibliography}
\end{document}